\documentclass[showpacs,twocolumn,floatfix]{revtex4}
\usepackage{graphicx}
\usepackage{dcolumn}
\usepackage{bm}
\usepackage{amssymb}
\usepackage{amsmath}
\usepackage{epsf}
\usepackage{subfigure}
\usepackage{epstopdf}%
\usepackage{amsfonts}
\usepackage{color}

\begin{document}

\title{Pulse propagation in decorated granular chains: An analytical
approach}
\author{Upendra Harbola$^1$, Alexandre Rosas$^2$, Aldo H. Romero
$^3$, Massimiliano Esposito$^{1,4}$ and Katja Lindenberg$^1$}
\affiliation{${}^1$Department of Chemistry and Biochemistry, and
BioCircuits Institute, University of California San Diego, La
Jolla, California 92093-0340, USA\\
${}^2$Departamento de F\'{\i}sica, CCEN, Universidade Federal da Para\'{\i}ba, Caixa Postal
5008, 58059-900, Jo\~ao Pessoa, Brazil\\
${}^3$Cinvestav-Quer\'etaro, Libramiento Norponiente 200, 76230, Fracc. Real de
Juriquilla, Quer\'etaro, Quer\'etaro, M\'exico\\
${}^{4}$Center for Nonlinear Phenomena and Complex Systems,
Universit\'e Libre de Bruxelles, Code Postal 231, Campus Plaine, B-1050
Brussels, Belgium\\}

\date{\today}

\begin{abstract}
We study pulse propagation in one-dimensional chains of spherical
granules decorated with small grains placed between large granules. The
effect of the small granules
can be captured by replacing the decorated chains by undecorated chains of large
granules of appropriately renormalized mass and effective interaction between
the large
granules. This allows us to obtain simple analytic
expressions for the pulse propagation properties using a generalization of the
binary collision approximation introduced in our earlier work [Phys. Rev. E
in print (2009); Phys. Rev. E {\bf 69}, 037601 (2004)].
\end{abstract}


\maketitle

\section{Introduction}
\label{introduction}

It is well known that an initial impulse applied at one end of a granular chain
in the absence of gaps or precompression can result in
solitary waves propagating through the medium \cite{nesterenko,nesterenko-1}. Apart from addressing fundamental problems of pulse
dynamics in the presence of highly nonlinear interactions, this scenario also
has direct
practical application, e.g., in designing shock absorbers
\cite{DarioPRL06,HongPRL05}, sound scramblers \cite{VitaliPRL05,DaraioPRE05} and
actuating devices \cite{KhatriSIMP08}. Thus, in the recent past pulse
propagation in one-dimensional (1D) chains of granules has been studied
extensively both
theoretically and experimentally
\cite{nesterenko-book,nakagawa,application,alexandre,alexandremono,jean,
wangPRE07,rosasPRL07,rosasPRE08,sen-review,costePRE97,nesterenkoJP94,
sokolowEPL07,wu02,robertPRE05,sokolowAPL05,senPHYA01,meloPRE06,jobGM07}.

Several configurations and parametrizations of the 1D chain have been used to
study pulse propagation in the face of
dissipation~\cite{application,alexandre,rosasPRL07,rosasPRE08} and of
polydispersity in the structure and mass of the
granules~\cite{nakagawa,wangPRE07,sen-review,robertPRL06,wu02,robertPRE05,
sokolowAPL05,meloPRE06,jobGM07,senPHYA01}.
In particular, polydispersity is frequently introduced in a regular fashion such
as in tapered chains
(TCs), in which the size and/or mass of successive granules
systematically decreases or
increases~\cite{nakagawa,wu02,robertPRE05,sokolowAPL05,meloPRE06,jobGM07,
senPHYA01,robertPRL06}.
Polydispersity is also introduced by distributing masses
randomly~\cite{SokolowSen07,ChenWang07}, by ``decorating" chains with
small masses placed regularly or randomly among larger
masses~\cite{robertPRL06,sen-review}, and by optimizing grain distribution
\cite{DaraioMAMS09}
for particular purposes. In a recent study
\cite{sen-review,robertPRL06} it was shown that the shock absorbing
properties of a granular chain can be enhanced quite dramatically by decorating the chain with smaller granules.
To our knowledge, however, most studies on decorated chains have been
numerical and hence restricted to specific parameter values.
One exception is the early work of Nesterenko and Laizardi~\cite{twomasses}
based on a continuum approximation, which is difficult to implement for chains
with rapid spatial variations such as are inherently present in decorated
chains. It is clearly
desirable to devise new theoretical
means to obtain analytic
results for these systems. This is the goal of the present work.

In a recent paper~\cite{PRE09} we analyzed pulse propagation in a variety of
undecorated TCs.
Using a binary collision approximation, we presented analytical expressions for
various quantities that characterize
propagation of the pulse or pulse front in 1D TCs.
However, an essential element of the success of this approximation is that a
given pair of granules collides only once, that is that there be no rebounds.
This approach is therefore not
directly applicable to decorated chains, where one (or more) small/light
granules between two large/heavy ones will rattle back and forth. A
similar conclusion about the inapplicability of a two-body
collision approximation (different from ours) to decorated
chains was also reached in Ref.~\cite{robertPRL06}.
On the other hand, the simplicity offered by the binary collision approximation
is too tempting to discard it altogether. Instead, we have succeeded in
extending
this theoretical scheme to decorated chains.

Thus, in this work we present such a scheme that allows us to make use of the
binary collision approximation for decorated chains in which single small/light
granules are placed between large/heavy ones. The case of more than one small
granule placed between large ones is not yet part of our scheme, nor is it
in most decorated configurations that have been considered (numerically) in
the literature. Such a generalization is however entirely feasible. Our approach
is based on a mapping of the decorated chain
onto an effective undecorated chain. The effect of small decorating granules is
included
through a modified interaction between the bigger granules. This also requires
the renormalization of the mass of the bigger granules. We present a systematic
way to carry out this renormalization. The binary approximation is then used on
the effective chain, which allows us to obtain analytical results for
the quantities that characterize the pulse front propagation in these chains. We
compare the analytical results with exact numerical results for
two different decorated chains: simple decorated chains, where the sizes of
the bigger particles are all the same, and decorated linear forward TCs,
where the sizes of the
bigger particles decrease linearly along the chain.

In Sec.~\ref{model} we introduce the general granular chain model in terms of rescaled (dimensionless) variables.
In Sec.~\ref{decorated-chains} we introduce decorated chains and present an
effective description in terms of the
renormalized interaction and masses. We then apply the binary
collision approximation in Sec.~\ref{binary} and obtain analytical results for
pulse properties. The analytic results are compared to
numerical integration results in Sec.~\ref{numerics}.
In Sec.~\ref{conclusion}
we provide a final summary and some comments about ongoing work.

\section{The model}
\label{model}

We consider chains of granules all made of the same material of density $\rho$.
When neighboring granules collide, they repel each other according to
the power law potential
\begin{equation}
 \label{hertz-1}
 V=\frac{a}{n} r_k^\prime |y_k-y_{k+1}|^n.
\end{equation}
Here $y_k$ is the displacement of granule $k$ from its position at the beginning of the collision,
and $a$ is a constant determined by Young's modulus and Poisson's ratio~\cite{landau,hertz}.
The exponent $n$ is $5/2$
for spheres (Hertz potential)~\cite{hertz}. We have
defined
\begin{equation}
  r_k^\prime = \left(\frac{2 R_k^\prime R_{k+1}^\prime}{R_k^\prime +R_{k+1}^\prime }\right)^{1/2},
\end{equation}
where $R_k^\prime$ is the principal
radius of curvature of the surface of granule $k$ at the point of contact.
When the granules do not overlap, the potential is zero.
The equation of motion for the $k$th granule is
\begin{eqnarray}
 \label{EOM-1}
 M_k\frac{d^2y_k}{d{\tau}^2}&=& {a}{r}_{k-1}^\prime(y_{k-1}-y_{k})^{n-1}\theta(y_{k-1}-y_k)\nonumber\\
&-& {a}{r}_{k}^\prime(y_k-y_{k+1})^{n-1}\theta(y_k-y_{k+1}),
\end{eqnarray}
where $M_k=(4/3)\pi \rho (R_k^\prime)^3$.  The Heaviside function
$\theta(y)$ ensures that the elastic interaction between grains only exists if
they are in contact. For the leftmost (rightmost) granule in the chain, only
the second (first) term on the right hand side of Eq.~(\ref{EOM-1}) is present.
Initially the granules are placed along a line so that they
just touch their neighbors in their equilibrium positions (no
precompression), and
all but the leftmost particle are at rest. The initial velocity of
the leftmost particle ($k=1$) is $V_1$ (the impulse).
We define the dimensionless quantity
\begin{equation}
\alpha\equiv \left[ \frac{M_1V_1^2}{a \left(R_1^\prime\right)^{n+1/2}}\right]
\end{equation}
and the rescaled quantities $x_k$, $t$, $m_k$, and $R_k$ via the relations
\begin{eqnarray}
y_k = R_1^\prime\alpha^{1/n} x_k, &\qquad&
\tau = \frac{R_1^\prime}{V_1} \alpha^{1/n} t, \nonumber\\
R_k^\prime = R_1^\prime R_k, &\qquad& M_k=M_1 m_k.
\end{eqnarray}
Equation~(\ref{EOM-1}) can then be rewritten as
\begin{eqnarray}
  m_k \ddot{x}_k &=&
 r_{k-1}(x_{k-1} - x_k)^{n-1} \theta (x_{k-1} -
x_k) \nonumber\\
&&-  r_k (x_k - x_{k+1})^{n-1} \theta (x_k -
x_{k+1}),
\label{eq:motion_rescaled}
\end{eqnarray}
where a dot denotes a derivative with respect to $t$, and
\begin{equation}
 \label{r-prime}
r_k= \left(\frac{2 R_kR_{k+1}}{R_k+R_{k+1}}\right)^{1/2}.
\end{equation}
The rescaled initial velocity is unity, i.e., $v_1(t=0)=1$.
The velocity of the $k$-th granule in unscaled variables is simply $V_1$ times its velocity in the
scaled variables.

\section{decorated chains and the effective dynamics}
\label{decorated-chains}

There are several ways to construct a decorated chain. We shall consider
decorated chains where small granules
(all of the same size) alternate with larger granules.
Our goal is to arrive at an analytically manageable effective description of a
decorated chain.  To be useful, such a description must also be accurate in
that it should capture (hopefully quantitatively) certain properties of the
pulse dynamics. In particular, we attempt to capture the time dependence of the
velocity profiles of the granules as the pulse moves along the chain, and the
time
that it takes a pulse to travel along the chain. We also try to characterize
the frequency of oscillation of the small granules that separate each pair of
large granules.

The frequency of oscillation of the small granules can be calculated by
considering just a threesome
of grains, two large ones with a small one in between.  The analysis of a
three-grain chain is also a convenient way to detail the mathematical analysis
that is then repeated for a longer system.  The three-grain chain is
analyzed in Sec.~\ref{threegrainchain}. On the other hand, to characterize the
large granules that will constitute the elements of our effective chain, the
shortest unit that must be analyzed is one of five granules in the sequence
large-small-large-small-large.  This is the minimal chain that starts with a
large granule (as do our long chains), and has an interior granule with small
granules on either side (whose effects will be captured in the renormalized
mass and renormalized interactions). The five-grain chain is analyzed in
Sec.~\ref{long-chain}.  Our effective chain of arbitrary length will then
consist of all large granules, two end ones identified with the two end granules
in the five-grain sequence, and any number of interior granules identified with
the interior large granule in the five-granule sequence.

\subsection{Three-grain chain}
\label{threegrainchain}

We first consider a chain of
three granules labeled $k-1$, $k$ and $k+1$, granule $k$ being the
small one. As noted above, this analysis will tell us something about the
frequency of oscillation of the small granule, and it will also allow us to lay
out the approximation scheme to be used in longer chains.
The dynamics of the three granules is governed by the equations of motion
\begin{subequations}
\begin{eqnarray}
\label{eqa}
 \ddot{x}_{k-1} &=&  -{\mathcal R}(x_{k-1} - x_k)^{n-1} \theta (x_{k-1}
-x_k)\\
\label{eqb}
 \ddot{x}_{k+1} &=&  {\mathcal R}(x_k - x_{k+1})^{n-1} \theta (x_k
-x_{k+1})\\
\label{eqc}
 m \ddot{x}_k &=& {\mathcal R}(x_{k-1} - x_k)^{n-1} \theta (x_{k-1}
-x_k)\nonumber\\
                 &&- {\mathcal R}(x_k - x_{k+1})^{n-1} \theta (x_k -x_{k+1}).
\end{eqnarray}
\end{subequations}
To convey the method most clearly, we have here taken the masses $m_{k\pm 1}$ of
the larger particles to be equal to each other and equal to unity in rescaled
variables; $m$ is the mass of the small granule. The effective elastic
 constant ${\mathcal R}$ is
given by Eq.~(\ref{r-prime}) for a large granule of unit radius adjacent to a
small one of radius $r$, that is,
\begin{equation}
{\mathcal R}=\left(\frac{2r}{1+r}\right)^{1/2}.
\label{radiusren}
\end{equation}
The
generalization to tapered
chains is implemented later.

When granule $k-1$ is given an initial impulse, the energy is transported
to granule $k$, which in turn transfers it to the
last granule $k+1$. Since granule $k$ is small, it gets compressed between the
two large granules and oscillates between them during
the energy transfer process. The frequency of these oscillations can be
expressed
in terms of the maximum force that the granule
$k$ experiences from granule $k+1$ or $k-1$. We follow Melo $et$
$al.$~\cite{melo} and write,
\begin{eqnarray}
 \label{displacement-1}
x_k(t) = \bar{x}_k(t)+A \mbox{sin}(\omega t+\phi)
\end{eqnarray}
Here $A$, $\omega$ and $\phi$ depend on the size ratio between the small and
large granules and may also depend on the time. We assume that these time
dependences are negligibly weak
during the time of energy transfer from $k-1$ to $k+1$. The first term
$\bar{x}_k$ represents
the average motion around which the particle executes oscillations of small
amplitude $A$ with frequency $\omega$. Substituting Eq.~(\ref{displacement-1})
in Eq.~(\ref{eqc}), and assuming that the amplitude $A$ is much smaller
than $\bar{x}_k-x_{k+1}$ and $x_{k-1}-\bar{x}_k$, we expand in $A$ and
from the lowest two orders we obtain
\begin{subequations}
\begin{eqnarray}
 \label{eq-2}
m\ddot{\bar{x}}_k  &\approx&
{\mathcal R}(x_{k-1}-\bar{x}_k)^{n-1}\theta(x_{k-1}-\bar{x}_k) \nonumber \\
&&-{\mathcal R}(\bar{x}_k-x_{k+1})^{n-1}\theta(\bar{x}_k-x_{k+1})\label{a}\\
m\omega^2 &\approx&
{\mathcal R}(n-1)\left[(x_{k-1}-\bar{x}_k)^{n-2}\theta
(x_{k-1}-\bar{x}_k)\right.\nonumber\\
&& \left. +(\bar{x}_k-x_{k+1})^{n-2}\theta (\bar{x}_k-x_{k+1})\right]. \label{b}
\end{eqnarray}
\end{subequations}
In the following, for economy of notation we do not the write $\theta$-functions
explicitly, but their presence restricting the interactions to compressions
should be kept in mind.

The small granule oscillates with maximum frequency around the ``equilibrium''
position, defined as the position at which the forces $\bar{F}$ obtained using
the average displacement of the smaller granule, cancel
each other on the two sides, i.e., where $\ddot{\bar{x}}_k=0$. Setting
\begin{equation}
 \bar{F} = {\mathcal R}(x_{k-1}-\bar{x}_k)^{n-1} = {\mathcal
R}(\bar{x}_k-x_{k+1})^{n-1}
\end{equation}
[cf. Eq.~(\ref{a})], and using Eq.~(\ref{b}), we can
write for the maximum frequency $f=\omega/2\pi$,
\begin{eqnarray}
 \label{eq-3}
f \approx
\frac{1}{2\pi}\sqrt{\frac{2(n-1)}{m}}{\mathcal
R}^{\frac{1}{2(n-1)}}\bar{F}^{\frac{ n-2 } { 2(n-1)}}.
\end{eqnarray}
This is the theoretical prediction for the frequency of the small granule that
we will later compare with numerical simulation results once we are able to
calculate the value of the average force $\bar{F}$ using our binary collision
approximation.

At the time when the small particle executes oscillations with maximum
frequency, we have $\ddot{\bar{x}}_k=0$. From Eq.~(\ref{a}), this implies that
\begin{eqnarray}
\label{eq-1aa}
\bar{x}_k=\frac{1}{2}(x_{k+1}+x_{k-1}).
\end{eqnarray}
Also, note that from Eqs.~(\ref{eqa})-(\ref{eqc})
\begin{eqnarray}
 \label{eq-1aaaa}
m\ddot{x}_{k}+\ddot{x}_{k-1}+\ddot{x}_{k+1}=0.
\end{eqnarray}
Equation~(\ref{eq-1aaaa}) is the statement that the net force on the system is
zero.
Substituting Eq.~(\ref{displacement-1}) in (\ref{eq-1aaaa}), using
Eq.~(\ref{eq-1aa}), and ignoring the small oscillatory part,
we get
$\left(\frac{m}{2}+1\right)(\ddot{x}_{k-1}+\ddot{x}_{k+1})=0$,
which describes a system of two granules each with a renormalized mass
\begin{equation}
\mu=1+\frac{m}{2}.
\label{massren}
\end{equation}
Replacing $x_k$ in Eqs.~(\ref{eqa}) and (\ref{eqb}) by the form assumed
in Eq.~(\ref{displacement-1}) and using Eq.~(\ref{eq-1aa}) we obtain
\begin{eqnarray}
 \label{eq-4}
\ddot{x}_{k-1} &\approx&
-\frac{\mathcal R}{2^{n-1}\mu}(x_{k-1}-x_{k+1})^{n-1}\nonumber\\
&& + \frac{m A}{2 \mu}\omega^2 \mbox{sin}{(\omega t+\phi)}\nonumber\\
\ddot{x}_{k+1} &\approx&
\frac{\mathcal R}{2^{n-1}\mu}(x_{k-1}-x_{k+1})^{n-1}\nonumber\\
&& - \frac{m A}{2 \mu}\omega^2 \mbox{sin}{(\omega t+\phi)}.
\end{eqnarray}
Since the oscillation amplitude $A$ is small, we neglect the oscillatory terms
in Eq.~(\ref{eq-4}) and simply write
\begin{eqnarray}
\mu\ddot{x}_{k-1} &\approx&
-\frac{\mathcal R}{2^{n-1}}(x_{k-1}-x_{k+1})^{n-1}\nonumber\\
\mu\ddot{x}_{k+1} &\approx&
\frac{\mathcal R}
{2^{n-1}}(x_{k-1}-x_{k+1})^{n-1}.
\label{eq-5}
\end{eqnarray}

Equations~(\ref{eq-5}) define the effective interaction potential $V_{eff}$
between granules $k-1$ and $k+1$.
Thus we can treat the system of two identical granules decorated with a smaller
granule as
an effective system of two granules
(now relabeled as $k$ and $k+1$) with effective masses $\mu$
interacting with an effective potential given by
\begin{eqnarray}
 \label{effective-potential}
V_{eff} = \frac{\mathcal R}{n2^{n-1}}(x_k-x_{k+1})^n.
\end{eqnarray}
Thus the overall effect of an intermediate granule is to reduce the interaction
strength between granules $k-1$ and $k+1$
(in the original chain) and increase their masses. This shows that the pulse
will be wider as
compared to the monodisperse case.
The result (\ref{effective-potential}) remains valid as long as the oscillation amplitude of the small granule is small
enough so that only the lowest order terms in the expansion, Eqs.~(\ref{a}) and
(\ref{b}), are important. We shall see that this puts an upper limit
on the size of the smaller granule.

\subsection{Long decorated chain}
\label{long-chain}

Using the three-grain chain to study a tapered chain leads to a certain
ambiguity because when there are two large granules of distinct masses around a
small one, the renormalized masses will be different depending on the pairing.
Thus, the renormalized mass of the $k$th granule would have different
values when considering the interaction between granules $k-1$ and $k$ or
between $k$ and $k+1$. This ambiguity can be suppressed if we consider a chain
of five granules centered on granule $k$, labeled from
$k-2$ to $k+2$. Granules $k-1$ and $k+1$ are small ones
of radius $r$. The radius of large granule $i=k,k\pm 2$ is $R_{i}>r$.
Moreover, this approach is useful even for a uniform decorated chain because it
naturally shows the distinction between a granule inside the chain and a
granule on either border.
The dynamics of this chain of granules is governed by the set of equations
\begin{eqnarray}
m_{k-2}\ddot{x}_{k-2} &=&  -r_{k-2}(x_{k-2} - x_{k-1})^{n-1} \theta (x_{k-2}
-x_{k-1})\nonumber\\
m_{i}\ddot{x}_{i} &=&  r_{i-1}(x_{i-1} - x_{i})^{n-1} \theta (x_{i-1} -x_i)\nonumber\\
                          -&&\hspace*{-0.2in}r_{i}(x_{i} - x_{i+1})^{n-1} \theta
(x_i -x_{i+1}), ~~ i=k, k\pm 1,\nonumber\\
m_{k+2}\ddot{x}_{k+2} &=&  r_{k+1}(x_{k+1} - x_{k+2})^{n-1} \theta (x_{k+1}
-x_{k+2}).
\nonumber\\
\label{lg-1}
\end{eqnarray}
For the smaller granules, as was done in Eq.~(\ref{displacement-1}), we again
assume a separation into an average displacement and an oscillatory
contribution about the average.
Following the same steps that led from Eq.~(\ref{displacement-1}) to
Eq.~(\ref{eq-1aa}), we now obtain for the smaller granules
\begin{eqnarray}
 \label{lg-2}
\bar{x}_{i} = \frac{x_{i-1}+\alpha_{i}x_{i+1}}{1+\alpha_{i}}
\end{eqnarray}
where $i=k-1,k+1$, and
\begin{eqnarray}
 \label{lg-3}
\alpha_i= \left(\frac{r_i}{r_{i-1}}\right)^{\frac{1}{(n-1)}}.
\end{eqnarray}

Using Eq.~(\ref{lg-2}) in Eq.~(\ref{lg-1}), the pulse propagation in a decorated
chain of five granules can be described by pulse propagation along an
effective chain of three large granules, the ``left'' ($l$), the ``middle''
($m$), and the ``right'' ($r$),
of renormalized masses
\begin{subequations}
\begin{eqnarray}
 \label{lg-4}
\label{ma}
\mu_{m} &=& m_k+\frac{m}{1+\left(\frac{r_{k-2}}{r_{k-1}}\right)^{\frac{1}{n-1}}}
+
\frac{m}{1+\left(\frac{r_{k+1}}{r_k}\right)^{\frac{1}{n-1}}} \nonumber \\ \\
\label{mb}
\mu_{l} &=& m_{k-2} +
\frac{m}{1+\left(\frac{r_{k-1}}{r_{k-2}}\right)^{\frac{1}{n-1}}}, \\ \nonumber\\
\label{mc}
\mu_{r} &=& m_{k+2} +
\frac{m}{1+\left(\frac{r_{k}}{r_{k+1}}\right)^{\frac{1}{n-1}}}.
\end{eqnarray}
\end{subequations}
Henceforth we relabel the three grains in the effective chain as $(l, m, r) \to
(k-1, k, k+1)$.
The effective interaction between the $k$th and $(k+1)$st granules
(relabeled) is given by
\begin{eqnarray}
 \label{lg-5}
V_{eff} = \frac{\zeta_k(n)}{n} (x_k-x_{k+1})^n,
\end{eqnarray}
where
\begin{eqnarray}
 \label{lg-6}
\zeta_k(n)=\frac{1}{\left[\left(\frac{\displaystyle
1}{\displaystyle r_k}\right)^{\frac{1}{n-1} }
+\left(\frac{\displaystyle
1}{\displaystyle r_{k+1}}\right)^{\frac{1}{n-1}}\right]^{n-1}},
\end{eqnarray}
and where $r_k$ is now given by
\begin{equation}
 r_k = \sqrt{\frac{2 R_k r}{R_k+r}}.
\end{equation}

A long decorated chain can now be represented by an effective chain of
large granules where the mass of the $k$th granule (except for those at the
edges of the chain) is $\mu_m$. Note that the effective mass $\mu_m$ is modified
from $m_k$ by the presence of the two
small granules on either side of it in the decorated chain.  The masses of the
granules at the edges of the long effective chain are $\mu_l$ and $\mu_r$ and
are modified from $m_{k\mp 2}$ by only a single small mass. The
interaction between the
$k$th and ($k+1$)st (relabeled) granules in the effective long chain is
given by Eq.~(\ref{lg-5}) with (\ref{lg-6}). Hence, as in the three-grain
chain, the interactions are weaker and the masses
larger in the effective chains than in the original decorated chain.
For the special case, $R_{k}=R_{k\pm 2}$ in the five-granule decorated chain,
Eqs.~(\ref{lg-2}) and (\ref{lg-5})
reduce to Eqs.~(\ref{eq-1aa}) and (\ref{effective-potential}), respectively.

\section{Binary collision approximation for effective chain}
\label{binary}

We now use the binary collision approximation to obtain analytic
expressions for the pulse propagating along a decorated chain and make the
connection with the effective description of Sec.~\ref{decorated-chains}. The
following analysis is based on
Ref.~\cite{PRE09}.

In the binary collision approximation we assume that the transfer of energy along the
chain occurs via a succession of two-particle collisions.
First, particle $k=1$
with velocity $v_1=1$ collides with initially stationary particle $k=2$, which
then acquires a velocity $v_2$
and collides with stationary particle $k=3$, and so on. Using energy and
momentum conservation and remembering that in the effective chain the granules
have effective mass $\mu_k$, it can easily be shown that
the velocity $v_k$ of the $k$th granule is given by
\begin{eqnarray}
 \label{binary-1}
v_k= \prod_{k^\prime=1}^{k-1}
\frac{2}{1+\frac{\displaystyle \mu_{k^\prime+1}}{\displaystyle \mu_{k^\prime}}}.
\end{eqnarray}

The coupled dynamics of any two interacting granules in our models can be
reduced to a single
particle dynamics. This is helpful toward obtaining analytical expressions for
some of the pulse properties, as we shall see below. For this purpose, we pick
granules $k$ and $k+1$ and introduce the difference variable
\begin{equation}
\label{variable}
  z_{k}=x_k-x_{k+1}.
\end{equation}
The equation of motion for the difference variable is obtained by subtracting
the equations of motion of the two granules during a collision,
\begin{eqnarray}
 \label{eom-11}
\ddot{x}_k &=& -\frac{\zeta_k(n)}{\mu_{k}}(x_k-x_{k+1})^{n-1}\nonumber\\
\ddot{x}_{k+1} &=& \frac{\zeta_k(n)}{\mu_{k+1}}(x_k-x_{k+1})^{n-1},
\end{eqnarray}
where
$\zeta_k(n)$ is defined in Eq.~(\ref{lg-6}).
Equation~(\ref{eom-11}) with (\ref{variable}) immediately leads to
\begin{equation}
\label{eq-z1}
\ddot{z}_{k}=-\frac{\zeta_k(n)}{{\mathcal M}_k} z_{k}^{n-1},
\end{equation}
where
\begin{equation}
{\mathcal M}_k=\frac{\mu_k\mu_{k+1}}{\mu_k + \mu_{k+1}}
\label{reducedmass}
\end{equation}
is the reduced mass of
the coupled system.

Equation~(\ref{eq-z1}) is the
equation of motion of a particle of effective mass ${\mathcal M}_k$ in the
potential $\zeta_k(n) z_k^n/n$ defined for $z_k\geq 0$.
The initial conditions are $\dot{z}_k(0)=v_k$ since the velocity of granule
$k+1$ is zero
before the collision, and $z_k(0)=0$ since there is no precompression.
The energy conservation condition
\begin{equation}
  \frac{1}{2} \dot{z}_k^2(t) +\frac{\zeta_k(n)}{n{\mathcal M}_k} z_k^n(t) =
\frac{1}{2} \dot{z}_k^2(0)
\label{max1}
\end{equation}
leads to
\begin{equation}
  \dot{z}_k(t) = \left(\dot{z}_k^2(0) -\frac{2\zeta_k(n)}{n
{\mathcal M}_k}z_k^n(t) \right) ^{1/2}.
\label{max}
\end{equation}

We say that the pulse arrives at granule $k$ when the velocity of granule $k$ surpasses that of
granule $k-1$, and that it moves on to granule $k+1$ when the velocity of the $(k+1)$st granule
surpasses that of the $k$th granule.  The residence time $T_k$ on granule $k$ is
the time that granule $k$ takes to transfer the pulse from $k-1$ to $k+1$, and is given by
\begin{equation}
\label{TK-eq}
T_k = \int_0^{z_k^{max}} \frac{dz_k}{\dot{z_k}}
 = \int_0^{z_k^{max}}\frac{dz_{k}}{\left(\dot{z}_k^2(0)
-\frac{\displaystyle 2\zeta_k(n)}{\displaystyle n {\mathcal M}_k}z_k^n(t)
\right) ^{1/2}},
\end{equation}
where $z_k^{max}$ is the compression when the velocities of particles $k$ and $k+1$ are equal
(which is also the maximum compression),
\begin{equation}
  z_k^{max} = \left( \frac{n {\mathcal M}_k}{2\zeta_k(n)} \dot{z}_k^2(0)\right)
^{1/n}.
\end{equation}
The integral can be performed exactly to yield
\begin{equation}
 \label{compress-time}
 T_k =
\sqrt{\pi}\left(\frac{n{\mathcal M}_k}{2\zeta_k(n)}\right)^{1/n}[\dot{z}_k(0)]^{
2/n-1 } \frac{\Gamma(1+1/n)}{\Gamma(1/n+1/2)}.
\end{equation}
Finally, the total time taken by the pulse to pass the $k$th granule is obtained by summing $T_k$,
\begin{equation}
t=\sum_{k^\prime=1}^k T_{k\prime}.
\label{totaltime}
\end{equation}
In the next section we compare these various quantities obtained from the binary
collision approximation implemented on the effective undecorated chains with
decorated chain numerical integration results.

At this point we can also calculate the average force that appears
in Eq.~(\ref{eq-3}).
When the two granules $k$ and $k+1$ collide, the maximum force between them corresponds to the maximum compression $z_k^{max}$. Thus the maximum force is
\begin{eqnarray}
 \label{max-force-general}
\bar{F} &=& \zeta_k(n) (z_{k}^{max})^{n-1}\nonumber\\
 &=& [\zeta_k(n)]^{1/n} \left(\frac{n{\mathcal M}_k
\dot{z}_k^2(0)}{2}\right)^{\frac{n-1}{n}}.
\end{eqnarray}
The explicit formulas for the three-grain configuration considered in
Sec.~\ref{threegrainchain} correspond to using Eq.~(\ref{reducedmass}) with
$\mu_k=\mu_{k+1} = \mu$ and hence ${\mathcal M}_k = \mu/2$, $\zeta_k(n) =
{\mathcal R} /2^{n-1}$, and $\dot{z}_k^2(0)=1$.  We obtain
\begin{eqnarray}
\label{max-force}
\bar{F} = \frac{n\mu}{4}\left(\frac{\mathcal R}{n2^{n-3}\mu}\right)^{1/n},
\end{eqnarray}
where ${\mathcal R}$ and $\mu$ are given in Eqs.~(\ref{radiusren}) and
(\ref{massren}), respectively.
Substituting this in
Eq.~(\ref{eq-3}), we obtain
\begin{eqnarray}
 \label{freq-1}
f \approx \frac{1}{2\pi}\sqrt{\frac{2(n-1)\mu}{m}}
\left(\frac{\mathcal R}{\mu}\right)^{1/n}
\left(\frac{n}{8}\right)^{\frac{n-2}{2n}}.
\end{eqnarray}
In the next section we compare this frequency with that obtained
from the numerical solution of the exact dynamical
equations (\ref{eqa})-(\ref{eqc}).

\section{Comparison with numerical results}
\label{numerics}

\begin{figure}[h]
\centering
\rotatebox{0}{\scalebox{.30}{\includegraphics{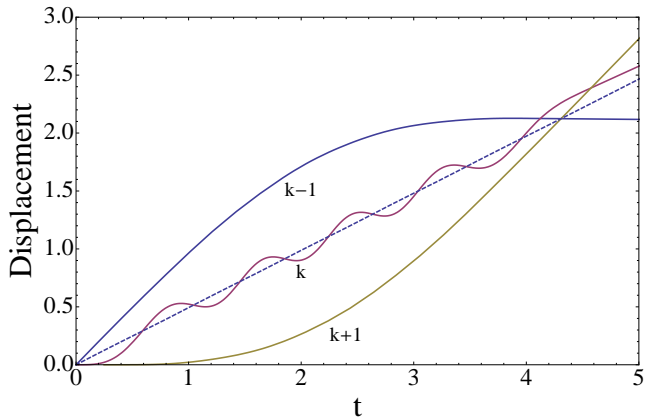}}}
%\rotatebox{0}{\scalebox{.30}{\includegraphics{Figure1new.eps}}}
\caption{(Color online) Solid lines: displacements of the three granules in a
three-granule
chain as a function of time.  Dashed line: average displacement of the small
middle granule.}
\label{figure0}
\end{figure}

While our binary collision approximation is valid for
$n>2$~\cite{alexandremono}, in our numerical comparisons we focus on spherical
granules. We start with the chain of three particles of
Sec.~\ref{threegrainchain}.  Our first figure, Fig.~\ref{figure0}, shows that
the important step of neglecting oscillatory terms as detailed in the
derivation of Eqs.~(\ref{a}), (\ref{b}) and (\ref{eq-5}) is
justified. The figure shows the displacement of the three granules as a
function of time.  The dashed curve shows the average displacement,
Eq.~(\ref{eq-1aa}), of the small granule.  The ``small amplitude'' in question
is the difference between the oscillatory curve and the dashed curve compared
to the differences between the dashed curve and the two large-granule
displacement curves.
%%%%%%%%%%%%%%%%%%%%%%%%%%%%%%%%%%%%%%%%%%%%%%%%%%%%%%%%%%%%%%%%%%%%%%%%

\begin{figure}[h]
\centering
\rotatebox{0}{\scalebox{.27}{\includegraphics{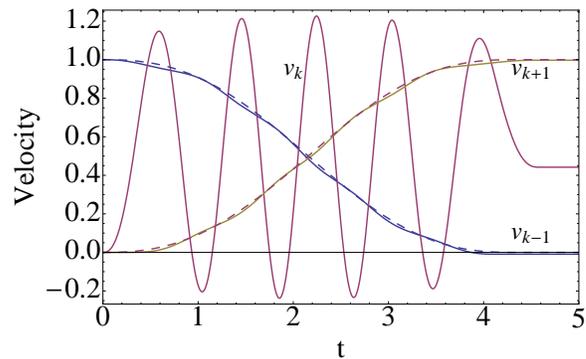}}}
%\rotatebox{0}{\scalebox{.30}{\includegraphics{Figure1new.eps}}}
\caption{(Color online) Typical velocity profiles of three granules (continuous
curves)
obtained from the numerical solution of the
exact dynamical equations. The small particle of radius $r=0.3$ oscillates with
almost constant frequency while the amplitude of oscillations
shows a weak time dependence. The dashed curves represent results from the
effective dynamics.}
\label{figure1}
\end{figure}

Figure~\ref{figure1} shows the velocity profile of the granules obtained from
the
solution of the exact dynamical
equations for the three granules. The small granule oscillates a number of times
before the chain disintegrates. In the same figure we also show a comparison
with the
results obtained from the effective dynamical equations~(\ref{eq-5}).
%%%%%%%%%%%%%%%%%%%%%%%%%%%%%%%%%%%%%%%%%%%%%%%%%%%%%%%%%%%%
%%%%%%%%%%%%%%%%%%%%%%%%%%%%%%%%%%%%%%%%%%%%%%%%%%%%%%%%%%%%
As we see from the figure, the behavior of the large particles is captured
very accurately by the effective interaction given by
Eq.~(\ref{effective-potential}).
We observe this quality of agreement as long as the radius of the
small granule is less than about $40\%$ of the larger ones. We return to this
point later in the paper. The effective
description improves as the size of the small granule decreases.

\begin{figure}[h]
\centering
\rotatebox{0}{\scalebox{.35}{\includegraphics{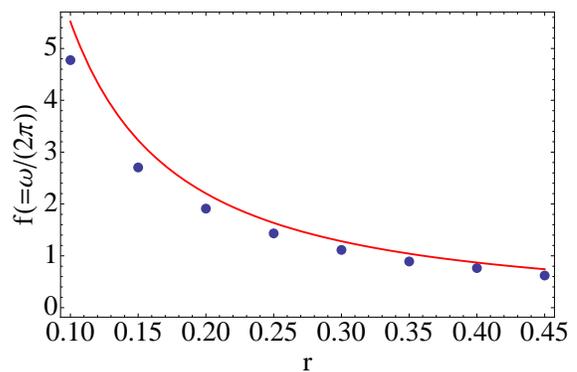}}}
%\rotatebox{0}{\scalebox{.30}{\includegraphics{Figure1new.eps}}}
\caption{(Color online) The dots represent the numerical results for the average
angular
frequency of oscillations of the small particle during the time of energy
transfer from
granule $k-1$ to $k+1$. The solid curve is the theoretical estimate based on the
binary collision approximation using the effective interaction.}
\label{figure2}
\end{figure}

Although it is not necessary for the characterization of the propagation of the
pulse, it is interesting to compare the theoretical [Eq.~(\ref{freq-1})] and
numerical results for
the oscillation frequency of the small particle as a function of its radius. The
theoretical result is shown as a solid curve in
Fig.~\ref{figure2}. The numerical results are obtained by fitting the
derivative of Eq.~(\ref{displacement-1}), $v(t) = 0.5 +\omega A \cos(\omega
t+\phi)$) (the particle velocity oscillates around $0.5$). The  resulting
frequencies are shown as dots in the figure. While the theoretical curve
appears closer to the numerical results for the larger granules, the percent
difference between the two is 14\% for the smallest granules and 20\% for the
largest granules shown in the figure.
%%%%%%%%%%%%%%%%%%%%%%%%%%%%%%%%%%%%%%%%%%%%%%%%%%%%%%%%%%%%
%%%%%%%%%%%%%%%%%%%%%%%%%%%%%%%%%%%%%%%%%%%%%%%%%%%%%%%%%%%%

\subsection{A simple decorated chain}

\begin{figure}[h]
\centering
\rotatebox{0}{\scalebox{.30}{\includegraphics{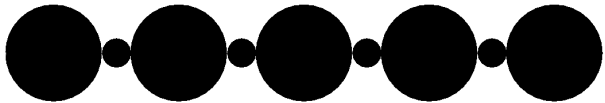}}}
%\rotatebox{0}{\scalebox{.25}{\includegraphics{simpleDC.9granules.eps}}}
%\rotatebox{0}{\scalebox{.30}{\includegraphics{Figure1new.eps}}}
\caption{Schematic of a simple decorated chain of nine granules.}
\label{schematic1}
\end{figure}

A simple decorated chain is one in which the large granules all have the same
radius.
For our numerical simulations of a simple decorated chain we consider
$N=50$ large granules, with each pair of large granules
separated by a small granule. Initially all the granules
are placed such that they barely touch their
nearest neighbors (no precompression). In all decorated chains, we consider
the first granule to be a large one. A schematic of a typical simple
decorated chain is shown in Fig.~\ref{schematic1}.
%%%%%%%%%%%%%%%%%%%%%%%%%%%%%%%%%%%%%%%%%%%%%%%%%%%%%%%%%%%%
%%%%%%%%%%%%%%%%%%%%%%%%%%%%%%%%%%%%%%%%%%%%%%%%%%%%%%%%%%%%%%%%%%%

\begin{figure}[h]
\centering
\rotatebox{0}{\scalebox{.30}{\includegraphics{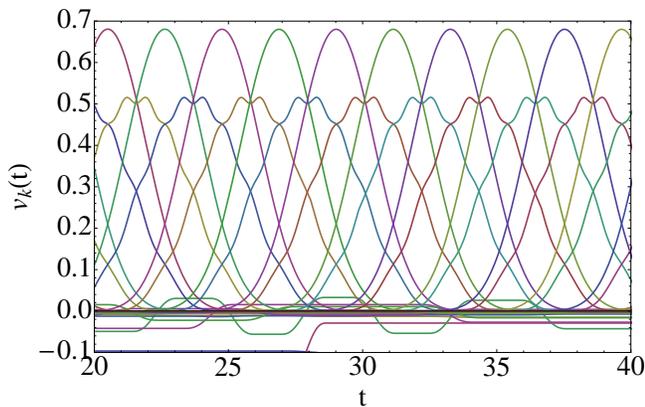}}}
%\rotatebox{0}{\scalebox{.30}{\includegraphics{Figure1new.eps}}}
\caption{(Color online) Velocity profile of granules in a simple decorated chain
with small
granules of radii $r=0.3$.}
\label{figure3}
\end{figure}

Figure~\ref{figure3} shows the
velocity
pulse propagating through the chain. The smooth higher velocity peaks are
associated with the large granules, while the lower peaks with oscillatory
contributions are those of the small granules. The oscillations in the
velocity profiles of the small particles are most
pronounced around their maximum velocity. This is due to the fact that the
forces on the two sides
of the small granule tend to cancel each other around that time.
%%%%%%%%%%%%%%%%%%%%%%%%%%%%%%%%%%%%%%%%%%%%%%%%%%%%%%%%%%%%
%%%%%%%%%%%%%%%%%%%%%%%%%%%%%%%%%%%%%%%%%%%%%%%%%%%%%%%%%%%%
Almost all the noisy behavior at the bottom of Fig ~\ref{figure3} is due to
the oscillations of the smaller granules. None of these oscillatory and noisy
contributions affect the smoothness of the
pulse front located on the larger
granules. We also note that the initial energy given to the left edge of the
chain
is transported mainly through the larger granules. The energy of the small
granules is negligible compared to that of the larger ones.

\begin{figure}[h]
\centering
\rotatebox{0}{\scalebox{.30}{\includegraphics{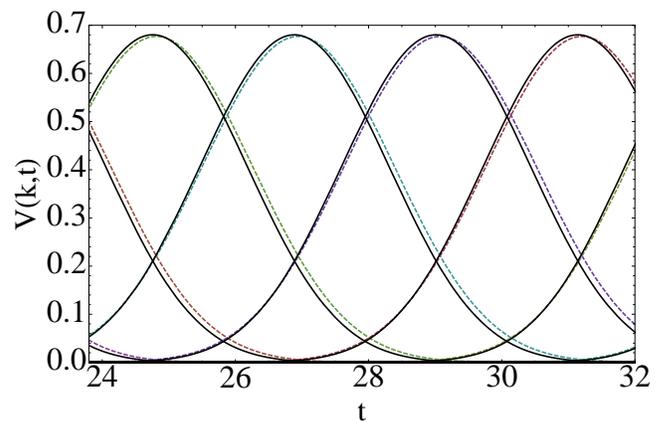}}}
%\rotatebox{0}{\scalebox{.30}{\includegraphics{Figure1new.eps}}}
\caption{(Color online) Exact results (solid curves) for the velocity profiles
of the larger
granules in a simple decorated chain for $r=0.3$. The dotted curves show results
from the solution of the effective dynamical equations.}
\label{figure4}
\end{figure}

\begin{figure}[h]
\centering
\rotatebox{0}{\scalebox{.80}{\includegraphics{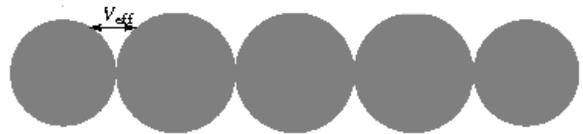}}
}
%\rotatebox{0}{\scalebox{.65}{\includegraphics{Effective-simpleDC.9granules.eps}
%}
%\rotatebox{0}{\scalebox{.30}{\includegraphics{Figure1new.eps}}}
\caption{Schematic of the effective undecorated chain associated with
Fig.~\ref{schematic1}. Each pair of granules interacts with the same pair
potential $V_{eff}$ but the end masses are smaller than the interior masses.}
\label{schematic2}
\end{figure}

In Fig.~\ref{figure4} we compare the exact dynamics of the larger granules with
the results obtained from the solution of the dynamical equations of the
associated effective undecorated chain.  We show only a small stretch in the
figure, but the agreement is typical of the entire chain. A schematic of
this chain is shown in
Fig.~\ref{schematic2}. The dynamical equations for the two edges of the
effective chain are given in Eq.~(\ref{eq-5}) with the associated effective
mass (\ref{massren}), or equivalently,
the effective mass (\ref{mb}) or (\ref{mc}) for equal large granules.  The
interior masses
satisfy the dynamical equation (with $\theta$-functions omitted)
\begin{equation}
\ddot{x}_k=\frac{{\cal R}}{2^{n-1}\mu}
[(x_{k-1}-x_k)^{n-1}-(x_k-x_{k-1})^{n-1}]
\end{equation}
with the effective
mass (\ref{ma}), which for the simple decorated chain reduces to
\begin{equation}
\mu_k=\mu=1+m.
\label{interiorrenmass}
\end{equation}
For clarity, we only show a section of the chain between $k=8$ and $k=24$ (here
$k$ counts only the larger granules). The results from the effective dynamics
are in
excellent agreement with the exact solution.
%%%%%%%%%%%%%%%%%%%%%%%%%%%%%%%%%%%%%%%%%%%%%%%%%%%%%%%%%%%%
%%%%%%%%%%%%%%%%%%%%%%%%%%%%%%%%%%%%%%%%%%%%%%%%%%%%%%%%%%%%
%%%%%%%%%%%%%%%%%%%%%%%%%%%%%%%%%%%%%%%%%%%%%%%%%%%%%%%%%%%%
%%%%%%%%%%%%%%%%%%%%%%%%%%%%%%%%%%%%%%%%%%%%%%%%%%%%%%%%%%%%%%%%%%%

\begin{figure}[h]
\centering
\rotatebox{0}{\scalebox{.3}{\includegraphics{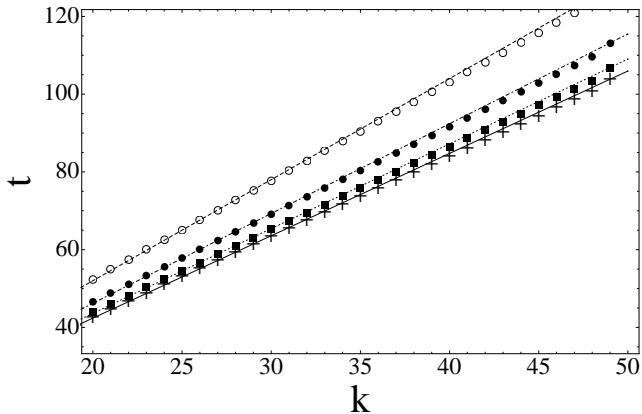}}}
%\rotatebox{0}{\scalebox{.30}{\includegraphics{Figure1new.eps}}}
\caption{Time $t$ taken by the pulse to reach the $k$th granule in a simple
decorated chain for $r=0.1$ (open circles), $r=0.2$ (filled circles), $r=0.3$
(squares), and $r=0.4 $ (plus signs). The
theoretical results are shown with dashed ($r=0.1$), dashed-dotted ($r=0.2$),
dotted ($r=0.3$), and continuous ($r=0.4$) lines.}
\label{figure6}
\end{figure}

For a simple decorated chain, $T_k$ is independent of $k$ (except for the edge
granules, which have a different effective mass) and is given by
\begin{eqnarray}
\label{TK}
T_k=
\sqrt{\pi}\left(\frac{n2^{n-3}\mu}{{\mathcal R}^{n-1}}\right)^{1/n}\frac{
\Gamma(1+1/n) }
{\Gamma(1/n+1/2)},
\end{eqnarray}
with the effective mass of Eq.~(\ref{interiorrenmass}).
The time $t$ defined in Eq.~(\ref{totaltime}) therefore varies linearly with
$k$. The slope of the line is determine by $T_k$.
In Fig.~\ref{figure6} we show numerical results for the time taken by the pulse
to reach the $k$th granule.
%%%%%%%%%%%%%%%%%%%%%%%%%%%%%%%%%%%%%%%%%%%%%%%%%%%%%%%%%%%%
%%%%%%%%%%%%%%%%%%%%%%%%%%%%%%%%%%%%%%%%%%%%%%%%%%%%%%%%%%%%
Note that $k$ here labels only the larger particles in the actual chain, and for
clarity we show results for the restricted range $20 < k <50$.
Various symbols represent different values of $r$. In the same figure we also
show the analytical result Eq.~(\ref{totaltime}) of the binary collision
approximation.
No fitting parameter are involved, and the agreement of the two results is
again gratifying. This shows that the effective binary theory
can give quantitatively reliable results for simple decorated chains.

\begin{figure}[h]
\centering
\rotatebox{0}{\scalebox{.30}{\includegraphics{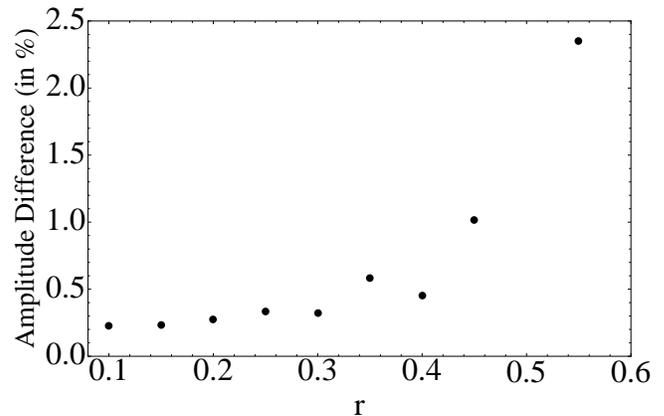}}}
%\rotatebox{0}{\scalebox{.30}{\includegraphics{Figure1new.eps}}}
\caption{Percent difference between the exact dynamics and the effective
dynamics as a function of the radius of the small granules. The
percent difference is an average calculated by taking the percent difference in
the pulse amplitude for each granule in the chain, summing over all granules,
and dividing by the total number of granules.}
\label{error}
\end{figure}

These quantitatively reliable results are obtained provided the radius of the
small granules is no larger than about $r \sim 0.4$.
In Fig.~\ref{error} we
show the percent difference in the
pulse amplitude calculated from the decorated dynamics and from the effective
undecorated one as a function of the radius of the small granules. The percent
difference is calculated by taking the percent difference in the pulse
amplitude for each granule in the chain, summing over all granules, and
dividing by the total number of granules.  The average percent difference
remains under
1\% up to about $r=0.4$ and beyond that rises sharply.
%%%%%%%%%%%%%%%%%%%%%%%%%%%%%%%%%%%%%%%%%%%%%%%%%%%%%%%%%%%%
%%%%%%%%%%%%%%%%%%%%%%%%%%%%%%%%%%%%%%%%%%%%%%%%%%%%%%%%%%%%%%%%%%%

\subsection{A tapered decorated chain}

\begin{figure}[h]
\centering
\rotatebox{0}{\scalebox{.30}{\includegraphics{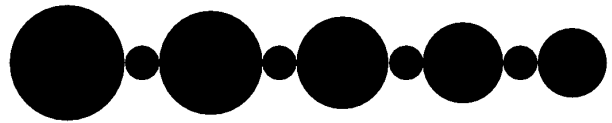}}}
%\rotatebox{0}{\scalebox{.25}{\includegraphics{grafix-TaperedDC.9granules.eps}}}
%\rotatebox{0}{\scalebox{.30}{\includegraphics{Figure1new.eps}}}
\caption{Schematic of a tapered decorated chain of nine granules.}
\label{schematic3}
\end{figure}

We next consider a linearly tapered chain decorated with small granules of
the same size. For numerical purposes, we again consider a tapered chain of
$N=50$ larger
granules. The radius of the $k$th granule is $R_k=1-S(k-1)$ and $S$ is the
tapering parameter~\cite{PRE09}. This chain is then decorated with the smaller
granules of
radius $r$. A schematic of a typical tapered decorated chain is shown in
Fig.~\ref{schematic3}.
%%%%%%%%%%%%%%%%%%%%%%%%%%%%%%%%%%%%%%%%%%%%%%%%%%%%%%%%%%%%
%%%%%%%%%%%%%%%%%%%%%%%%%%%%%%%%%%%%%%%%%%%%%%%%%%%%%%%%%%%%%%%%%%%

Our effective description is valid as long as the amplitudes of
oscillations of the smaller particles are small compared to their average
displacement.
This again occurs as long as the size of the small granules is roughly less than
$40\%$ of the size of the larger granules. The value of $S$ is therefore
restricted
by the length of the tapered chain.  This is the reason for choosing linear
rather than geometric tapering for this analysis~\cite{PRE09}, which is even
more restrictive.

\begin{figure}[h]
\centering
\rotatebox{0}{\scalebox{.3}{\includegraphics{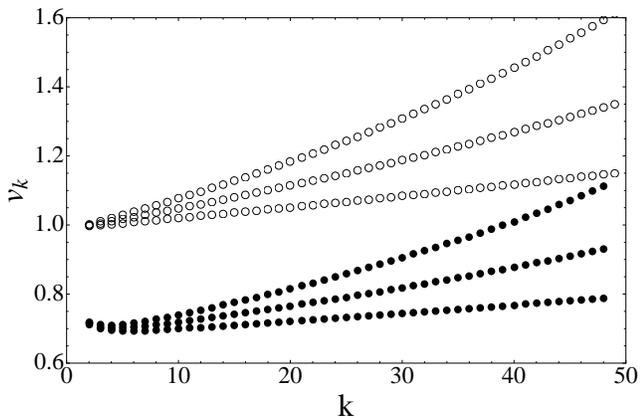}}}
%\rotatebox{0}{\scalebox{.30}{\includegraphics{Figure1new.eps}}}
\caption{Change in the pulse amplitude as a function of $k$. The filled and the
empty circles represent the exact numerical result and the binary
collision approximation
result, respectively. The three different data sets are for $S=0.002$,
$0.004$, $0.006$ from bottom to top. In all cases $r=0.3$.}
\label{figure7}
\end{figure}

In Fig.~\ref{figure7} we show the change in the pulse amplitude with the
granule number. As already noted in the case of undecorated tapered
chains~\cite{PRE09}, the absolute value of the pulse amplitude is
 %%%%%%%%%%%%%%%%%%%%%%%%%%%%%%%%%%%%%%%%%%%%%%%%%%%%%%%%%%%%
%%%%%%%%%%%%%%%%%%%%%%%%%%%%%%%%%%%%%%%%%%%%%%%%%%%%%%%%%%%%
not captured correctly by the binary approximation, but the rate of
change of the amplitude is close to that obtained from the exact results.

\begin{figure}[h]
\centering
\rotatebox{0}{\scalebox{.3}{\includegraphics{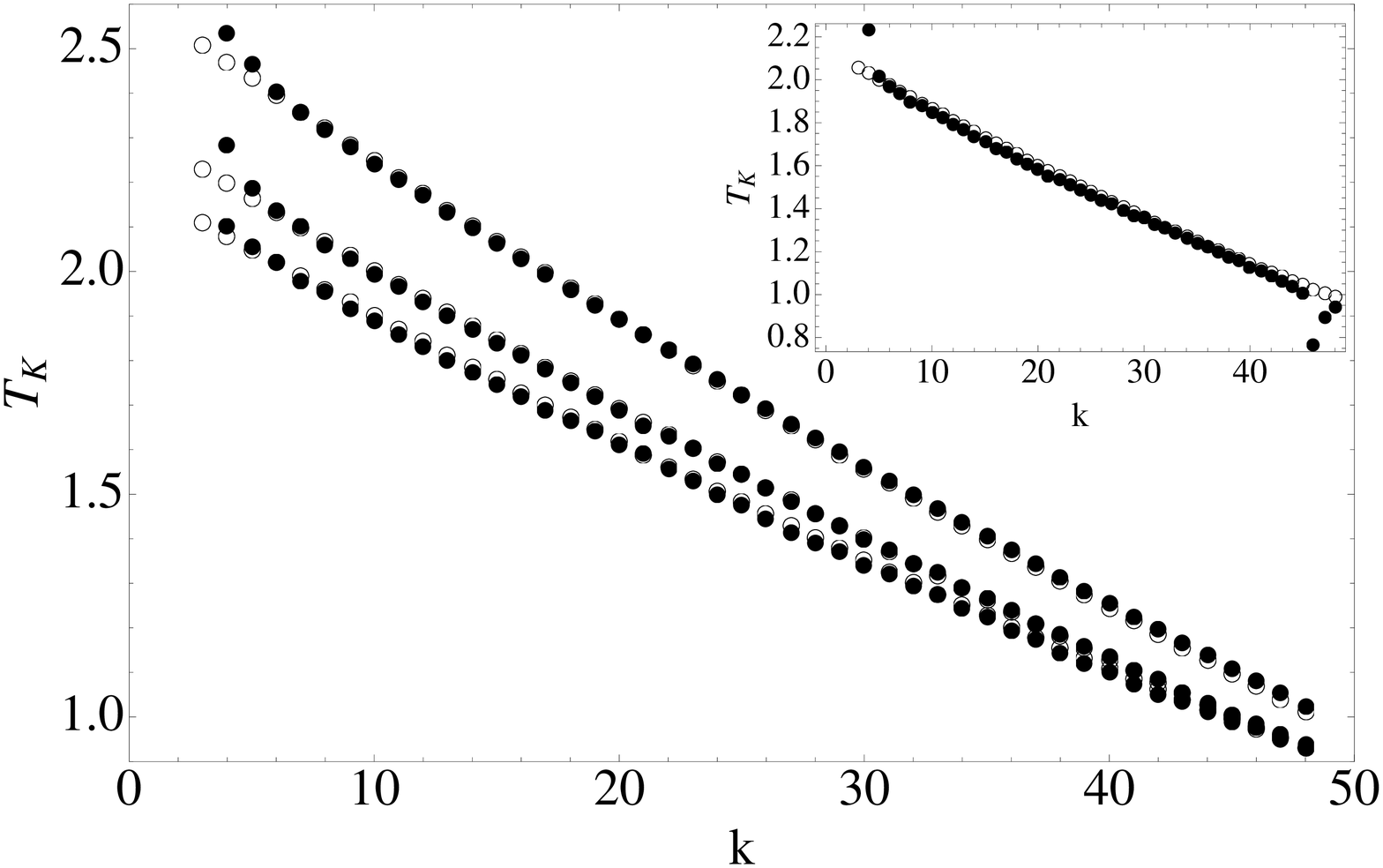}}}
%\rotatebox{0}{\scalebox{.30}{\includegraphics{Figure1new.eps}}}
\caption{Residence time $T_k$ of the pulse on the $k$th granule for
$r=0.1$, $0.2$, and $0.3$ from top to bottom, and $S=0.01$. Numerical and
analytical results are
represented
by the filled and open circles, respectively. The inset shows results for
$r=0.4$.}
\label{figure8}
\end{figure}

However, other characteristics of the pulse are extremely well captured by the
binary approximation. We next compare the binary prediction for the time spent
by each granule in the
pulse. The analytical result for the residence time of the pulse on granule $k$
is given by $T_k$ in Eq.~(\ref{TK-eq}).
In Fig.~\ref{figure8} we compare the results obtained from the exact dynamics
to those of the binary collision approximation for $r=0.1$, $0.2$, and $0.3$.
Excellent agreement is again obtained between the two results. The small initial
disagreement is due to an edge effect. The residence time of the pulse
decreases rapidly as the pulse propagates through the chain, as it does in
a non-decorated
forward tapered chain~\cite{PRE09}. Note that the residence time of the pulse
on granule $k$ decreases with increasing $r$ because the increase in the
effective mass is accompanied by an even greater increase in the effective
interaction bewteen granules. In the inset of Fig.~\ref{figure8}, we
show the results for $r=0.4$. Here the exact and approximate results disagree
for large $k$ values
because the size difference between the larger and the smaller granules
decreases to where the
description in terms of an effective chain breaks down.
%%%%%%%%%%%%%%%%%%%%%%%%%%%%%%%%%%%%%%%%%%%%%%%%%%%%%%%%%%%%
%%%%%%%%%%%%%%%%%%%%%%%%%%%%%%%%%%%%%%%%%%%%%%%%%%%%%%%%%%%%

\section{conclusions}
\label{conclusion}

We have studied pulse propagation in decorated chains, that is, chains in which
small and large granules alternate. We have presented a method to map the
actual chain into an effective undecorated chain. This effective description
allows us to obtain
analytical results using the binary collision approximation for quantities that characterize the pulse properties in decorated chains.
The effective description is extremely accurate as long as the size ratio of the
small granules to the large granules is less
than $\sim 0.40$. We have explicitly considered two different chains, simple
decorated chains and forward tapered decorated chains. For
simplicity, in order to obtain an effective description, we have assumed a fixed
size for the small granules.
However the method can be easily applied with little modifications to other more
complicated decorated chains where small granules are also allowed to change
size. The method is also implementable to decorated chains with more than one
large granule between the small ones.  A more extensive
generalization is
necessary to deal with decorated chains with more than one small granule
between large ones and thence to other size profiles.  This work is in progress.

\section*{Acknowledgments}
Acknowledgment is made to the Donors of the American Chemical Society Petroleum Research Fund for
partial support of this research (K.L. and U.H.).  A.R. acknowledges support from Pronex-CNPq-FAPESQ
and CNPq.  M. E. is supported by the FNRS Belgium (charg\'e de recherches) and
by the government of Luxembourg (Bourse de formation recherches). A. H. R.
acknowledges support by CONACyT Mexico under Projects J-59853-F and J-83247-F.

\end{document}